# The Special Theory of Relativity as Applied to the Born-Oppenheimer-Huang Approach

By

Michael Baer

**The Fritz Haber Center for Molecular Dynamics, The Hebrew University of Jerusalem, Jerusalem 91904, Israel**

[a]email: michaelb@fh.hujA.ac.il

# Abstract

In two recent publications ( Int. J. Quant. Chem. 114, 1645 (2014) and Molec. Phys. 114, 227 (2016)) it was shown that the Born -Hwang (BH) treatment of a molecular system perturbed by an external field yields a set of decoupled vectorial Wave Equations, just like in Electro-magnetism. This finding led us to declare on the existence of a new type of *Fields,* which were termed *Molecular Fields.* The fact that such fields exist implies that at the vicinity of *conical intersections* exist a mechanism that transforms a passing-by electric beam into a field which differs from the original electric field.

This situation is reminiscent of what is encountered in astronomy where Black Holes formed by massive stars may affect the nature of a near-by beam of light. Thus if the Non-Adiabatic-Coupling-Terms (NACT) with their singular points may affect the nature of such a beam (see the above two publications) then it would be interesting to know to what extend NACTs (and consequently also the BH equation) will be affected by the special theory of relativity as introduced by Dirac. Indeed while applying the Dirac approach we derived the relativistic affected NACTs as well as the corresponding BH equation.

## A. Introduction

For more than a decade we treat molecular systems exposed to external electric fields. Most of the ideas we had about this subject were published in a series of articles starting in 2003.[1-8] Step by step, this series developed testing new ideas until it reached its final publication just a year ago.[8] The main obstacle we encountered was how to include time, rigorously, while carrying out the Born-Huang (BH)[9a] expansion. The straightforward way is to treat the field-dressed case (namely, the time-dependent equation) just like one treats the field-free case, as originally proposed by Born and Oppenheimer (BO),[9b] namely by relating to the time independent eigen-value equation:[1,3a,10a,b,11]

$$\left(\mathbf{H}_{e0}(\mathbf{s}_e,\mathbf{s})-u_k(\mathbf{s})\right)|\varsigma_k(\mathbf{s}_e|\mathbf{s})\rangle=0; \quad k=\{1,N\} \tag{1}$$

Here $\mathbf{H}_{e0}(\mathbf{s}_e,\mathbf{s})$ is the (field-free) electronic Hamiltonian, $u_k(\mathbf{s})$ are the corresponding electronic eigen-values (which are also recognized as the adiabatic potential energy surfaces (PES)), $|\varsigma_k(\mathbf{s}_e|\mathbf{s})\rangle$; k={1,N} are the corresponding (adiabatic) electronic eigen-functions and finally $\mathbf{s}$ and $\mathbf{s}_e$ stand for collections of nuclear and electronic coordinates, respectively.

The other possibility is to include time within the treatment of the eigen-value problem. This change requires the inclusion of the corresponding time derivative so that Eq. (1) becomes: (1,2,3b)

$$i\hbar \frac{\partial}{\partial t}\left|\tilde{\zeta}_j(\mathbf{s}_e|\mathbf{s},t)\right\rangle = \mathbf{H}_e(\mathbf{s}_e,\mathbf{s},t)\left|\tilde{\zeta}_j(\mathbf{s}_e|\mathbf{s},t)\right\rangle; j=\{1,N\} \qquad (2)$$

Here $\mathbf{H}_e(\mathbf{s}_e,\mathbf{s},t)$ is the electronic Hamiltonian which also contains the external perturbation expressed in terms of an electric field interacting with the electrons, thus

$$\mathbf{H}_e(\mathbf{s}_e,\mathbf{s},t) = \mathbf{H}_{e0}(\mathbf{s}_e,\mathbf{s}) + \boldsymbol{\mu}_e(\mathbf{s}_e) \cdot \boldsymbol{E}(t) \qquad (3)$$

where $\boldsymbol{\mu}_e(\mathbf{s}_e)$ is the electronic dipole moment defined as $\boldsymbol{\mu}_e = e\sum \mathbf{s}_e$ and $\boldsymbol{E}(t)$ is the intensity of the electric (external) field.

The advantage of this approach is in its physical message, namely that the main effect of the external field is due to its interaction with the fast moving, light electrons and therefore it is expected that this effect has to be treated, from the *outset*, explicitly. At this stage it is important to emphasize that this novel approach was tested in a series of numerical studies, first applied to a model system [4-6] and later to a real system, namely $(H_2)^+$, for which photo-dissociation probabilities and kinetic energy distributions were calculated employing the two approaches.[12-15] The results were compared with other calculations as well as with experiments (see in particular in Ref. 15).

This approach was found, at later stages, to lead to Wave Equations reminiscent of the Maxwell-Lorentz (ML) Wave Equations that describe the Electromagnetic (EM) field.[16] However whereas the ML equations are based on the EM vector-potential (which results from the electric component **E** and the magnetic component **H** which satisfy Maxwell's equations) the present molecular Wave Equations have their origin in the BH vectorial Non-Adiabatic Coupling Terms (NACTs) :[10,11]

$$\boldsymbol{\tau}_{12} = \langle \zeta_1 | \nabla \zeta_2 \rangle \quad \text{and} \quad \boldsymbol{\tau}_{12}^{(2)} = \langle \zeta_1 | \nabla^2 \zeta_2 \rangle \tag{4}$$

where $\boldsymbol{\tau}_{12}(p,q)$ – a vector – is the NACT of the first order and $\boldsymbol{\tau}_{12}^{(2)}(p,q)$ – a scalar – is the NACT of the second order and $\nabla$ is the nuclear vectorial grad operator.. Here p and q are two (nuclear) Cartesian coordinates.

The Wave Equations are developed in two steps. In Ref. (7) the emphasize is treated the left-hand-side of the Wave Equation for a planar (consequently, a two-component system) defined in terms of two-states exposed to an external electric field. It was shown that the two corresponding time-dependent dressed Cartesian NACTs:

$$\tilde{\tau}_{p12} \quad and \quad \tilde{\tau}_{q12}$$

which are explicitly expressed in terms of various time dependent functions and the two time independent NACTs given in Eqs (4) were shown to fulfill the following Wave Equation:

$$\frac{\partial^2 \tilde{\tau}_{z12}}{\partial p^2} + \frac{\partial^2 \tilde{\tau}_{z12}}{\partial q^2} - \frac{1}{c^2}\frac{\partial^2 \tilde{\tau}_{z12}}{\partial t^2} = J_z(p,q,t); \quad z=p,q \qquad (5)$$

The efforts in Ref. (8) were centered on deriving the corresponding right-hand-side functions in Eqs. (5), namely, $J_p(p,q,t)$ and $J_q(p,q,t)$. These functions were first derived for the general case but then in order to simplify the expressions and make them more transparent they were worked out for the situation when (1) the intensity is of the external field is *high enough* and (2) when the duration time of the field is *short enough*. For these two assumptions are derived the two inhomogenuity terms:

$$J_z(p,q,t) = \cos\Theta \frac{\partial \tau_{12}^{(2)}}{\partial z} - \frac{1}{c^2}\tau_{z12}\frac{\partial^2(\cos\Theta)}{\partial t^2} \quad z=p,q \qquad (6)$$

where the function, $\Theta(p,q,t)$, is an integral over the external field intensity $\Phi_{e12}(p,q,t)$, thus:

$$\Theta(p,q,t) = -\frac{2}{\hbar}\int_0^t \Phi_{e12}(p,q,t')dt'; \qquad (7)$$

Here $\Phi_{e12}(p,q,t)$ stands for the product:

$$\Phi_{e12}(p,q,t) = \boldsymbol{\mu}_{12}(p,q) \cdot \boldsymbol{E}(t) \qquad (8)$$

where $\mu_{12}(p,q)$ is the *transition* dipole moment and $E(t)$ is the intensity of the electric (external) field (see also Eq. (3)).

The fact that the Born-Oppenheimer-Hwang (BOH) treatment of a molecular system perturbed by an external time dependent electric field yields a set of decoupled vectorial Wave Equations of the kind revealed for the EM case, *implies* that, here too, are encountered *Fields* that are termed: *Molecular Fields*.[7,8] Next, since the BO treatment [9b] tends to produce pairs of two intersecting/tangential *adiabatic* potential energy surfaces where each one forms a cone – see Fig.1 – surrounding the vicinity of their points of intersection (also known as a points of Conical Intersection (*ci*)) [3c,17-19] and since inside each of the cones are concentrated the NACTs which become more intense the closer is the *ci*-point, this new field seems to be a result of an interaction applied to the original electric beam (which, to start with, was trapped inside these two cones) and therefore may differ from the original electric field.

This situation, if it really exists, is somewhat reminiscent of what is encountered in astronomy/cosmology (still, to be taken with a grain of salt). Here numerous physicists considered the possibility that certain stars which are heavy enough may affect the path and the nature of a passing-by beam of light. These stars were termed by Wheeler as Black Holes [20] and are characterized by having a well defined metaphoric spacial "mantle" and a singularity point at its center. [20] These two features are similar of what happens in the present situation. Here the NACTs are wrappd-up by the adiabatic PESs and found to have a singularity, at its center. Such a system could be capable of affecting the nature of an electric beam trapped inside these cones as was discussed in references (7) and (8) and schematically shown in Fig.1.

In the present article we suggest to move one step further following the question to be asked: Since we deal with features reminiscent of cosmic phenomena can the just mentioned

Molecular Fields be affected by relativistic mass effects? In order to study this possibility we extend the BOH approach so that it includes the Dirac relativistic theory of the electron![21-22]

## B. Comments to the Born-Huang treatment

The starting point of the Born-Hwang (BH) treatment is the ordinary Schrödinger equation for the nuclear-electronic Hamiltonian **H** and the corresponding total wave-function, $\boldsymbol{\Psi}(\mathbf{s}_e,\mathbf{s})$ given in the form:

$$\mathbf{H}\boldsymbol{\Psi}(\mathbf{s}_e,\mathbf{s}) = E\boldsymbol{\Psi}(\mathbf{s}_e,\mathbf{s}) \qquad (9)$$

or more explicitly:

$$\left(-\frac{\hbar^2}{2M}\nabla^2 + \mathbf{H}_{e0}(\mathbf{s}_e|\mathbf{s})\right)\boldsymbol{\Psi}(\mathbf{s}_e,\mathbf{s}) = E\boldsymbol{\Psi}(\mathbf{s}_e,\mathbf{s}) \qquad (10)$$

where M is the mass of the system, E is the total energy, $\nabla$ is the nuclear gradient (vector) operator expressed in terms of mass-scaled nuclear coordinates, **s**. $\mathbf{H}_{e0}(\mathbf{s}_e|\mathbf{s})$ is the electronic Hamiltonian (which in the present case is assumed to be time-independent) introduce earlier (see Eq. (1))

Following the BH approach, the total wave function is expanded in terms of the previously introduced electronic eigen-functions (see Eq. (1)):

$$\Psi(\mathbf{s}_e,\mathbf{s}) = \sum_{k=1}^{N} \varsigma_k(\mathbf{s}_e\,|\,\mathbf{s})\psi_k(\mathbf{s}) \qquad (11)$$

where $\psi_k(\mathbf{s})$; k={1,N}, are the corresponding nuclear wave functions. Consequently the relevant Schrodinger equation for the $\psi_k(\mathbf{s})$'s can be shown to take the form [9a]

$$-\frac{\hbar^2}{2M}\nabla^2\psi_k + (u_k - E)\psi_k - \frac{\hbar^2}{2M}\sum_{n=1}^{N}\left(2\boldsymbol{\tau}_{kn}\cdot\nabla + \boldsymbol{\tau}_{kn}^{(2)}\right)\psi_n = 0;\ \ k=\{1,N\} \qquad (12)$$

where $\boldsymbol{\tau}(\mathbf{s})$, the NACT matrix of the first order and $\boldsymbol{\tau}^{(2)}(\mathbf{s})$, the NACT matrix of the second order were introduced in Eqs, (4) and $u_k$, the adiabatic PES was introduced in Eq. (1). For a system of real electronic wave functions $\boldsymbol{\tau}(\mathbf{s})$ is an anti-symmetric matrix.

Eq. (12) can also be written in a matrix form as follows: [3a,11]

$$-\frac{\hbar^2}{2M}\nabla^2\boldsymbol{\psi} + (\mathbf{u} - E)\boldsymbol{\psi} - \frac{\hbar^2}{2M}\left(2\boldsymbol{\tau}\cdot\nabla + \boldsymbol{\tau}^{(2)}\right)\boldsymbol{\psi} = \mathbf{0} \qquad (13a)$$

(where the dot designates the scalar product) or also [10]

$$-\frac{\hbar^2}{2M}(\nabla + \boldsymbol{\tau})^2\boldsymbol{\psi} + (\mathbf{u} - E)\boldsymbol{\psi} = \mathbf{0} \qquad (13b)$$

which, among other things, emphasizes the physical meaning of $\boldsymbol{\tau}(\mathbf{s})$.

## C. The Schrödinger-Dirac Equation

## C.1 The Dirac Electronic Hamiltonian

In Appendix I are derived the four coupled Schrödinger-Dirac Equations for a system of a single electron (see Eqs. I.11) or (I.12)):

$$-i\hbar \frac{\partial}{\partial t}|\zeta_1(\mathbf{s}_e|\mathbf{s},t)\rangle = ic\hbar \left[ \frac{\partial}{\partial s_{ex}}|\zeta_4(\mathbf{s}_e|\mathbf{s},t)\rangle - i\frac{\partial}{\partial s_{ey}}|\zeta_4(\mathbf{s}_e|\mathbf{s},t)\rangle \right.$$
$$\left. + \frac{\partial}{\partial s_{ez}}|\zeta_3(\mathbf{s}_e|\mathbf{s},t)\rangle \right] + \left(V(\mathbf{s}_e|\mathbf{s}) + mc^2\right)|\zeta_1(\mathbf{s}_e|\mathbf{s},t)\rangle \quad (14a)$$

$$-i\hbar \frac{\partial}{\partial t}|\zeta_2(\mathbf{s}_e|\mathbf{s},t)\rangle = ic\hbar \left[ \frac{\partial}{\partial s_{ex}}|\zeta_3(\mathbf{s}_e|\mathbf{s},t)\rangle + i\frac{\partial}{\partial s_{ey}}|\zeta_3(\mathbf{s}_e|\mathbf{s},t)\rangle \right.$$
$$\left. - \frac{\partial}{\partial s_{ez}}|\zeta_4(\mathbf{s}_e|\mathbf{s},t)\rangle \right] + \left(V(\mathbf{s}_e|\mathbf{s}) + mc^2\right)|\zeta_2(\mathbf{s}_e|\mathbf{s},t)\rangle \quad (14b)$$

$$-i\hbar\frac{\partial}{\partial t}|\zeta_3(\mathbf{s}_e|\mathbf{s},t)\rangle = ic\hbar\left[\frac{\partial}{\partial s_{ex}}|\zeta_2(\mathbf{s}_e|\mathbf{s},t)\rangle - i\frac{\partial}{\partial s_{ey}}|\zeta_2(\mathbf{s}_e|\mathbf{s},t)\rangle\right.$$
$$\left.+\frac{\partial}{\partial s_{ez}}|\zeta_1(\mathbf{s}_e|\mathbf{s},t)\rangle\right] + \left(V(\mathbf{s}_e|\mathbf{s}) - mc^2\right)|\zeta_3(\mathbf{s}_e|\mathbf{s},t)\rangle \quad (14c)$$

$$-i\hbar\frac{\partial}{\partial t}|\zeta_4(\mathbf{s}_e|\mathbf{s},t)\rangle = ic\hbar\left[\frac{\partial}{\partial s_{ex}}|\zeta_1(\mathbf{s}_e|\mathbf{s},t)\rangle + i\frac{\partial}{\partial s_{ey}}|\zeta_1(\mathbf{s}_e|\mathbf{s},t)\rangle\right.$$
$$\left.-\frac{\partial}{\partial s_{ez}}|\zeta_2(\mathbf{s}_e|\mathbf{s},t)\rangle\right] + \left(V(\mathbf{s}_e|\mathbf{s}) - mc^2\right)|\zeta_4(\mathbf{s}_e|\mathbf{s},t)\rangle \quad (14d)$$

Dirac studied the energy levels of the Hydrogen atom while employing two out of the four equations, namely, Eqs. (14a) and (14d) and ignoring the third term in each one of them thus leading to a planar electronic system).[21] We consider the same two coupled equations

$$-i\hbar\frac{\partial}{\partial t}|\zeta_1(\mathbf{s}_e|\mathbf{s},t)\rangle = \left(V(\mathbf{s}_e|\mathbf{s}) + mc^2\right)|\zeta_1(\mathbf{s}_e|\mathbf{s},t)\rangle +$$
$$ic\hbar\left(\frac{\partial}{\partial s_{ex}} - i\frac{\partial}{\partial s_{ey}}\right)|\zeta_4(\mathbf{s}_e|\mathbf{s},t)\rangle \quad (15a)$$

$$-i\hbar\frac{\partial}{\partial t}|\zeta_4(\mathbf{s}_e|\mathbf{s},t)\rangle = \left(V(\mathbf{s}_e|\mathbf{s}) - mc\right)^2|\zeta_4(\mathbf{s}_e|\mathbf{s},t)\rangle +$$
$$ic\hbar\left(\frac{\partial}{\partial s_{ex}} + i\frac{\partial}{\partial s_{ey}}\right)|\zeta_1(\mathbf{s}_e|\mathbf{s},t)\rangle \quad (15b)$$

As mentioned above we intend to study the planar system and therefore, in Appendix I, are expressed Eqs. (15) in term of (electronic) polar coordinates.

$$-i\hbar \frac{\partial}{\partial t}\left|\varsigma_k^{(1)}(r_e,\varphi_e \mid \mathbf{s},t)\right\rangle = \left(V(r_e,\varphi_e \mid \mathbf{s}) + mc^2\right)\left|\varsigma_k^{(1)}(r_e,\varphi_e \mid \mathbf{s},t)\right\rangle +$$
$$i\exp(-i\varphi_e)\left(\frac{\partial}{\partial r_e} - \frac{1}{r_e}\frac{\partial}{\partial \varphi_e}\right)\left|\varsigma_k^{(2)}(r_e,\varphi_e \mid \mathbf{s},t)\right\rangle \quad (16a)$$

$$-i\hbar \frac{\partial}{\partial t}\left|\varsigma_k^{(2)}(r_e,\varphi_e \mid \mathbf{s},t)\right\rangle = \left(V(r_e,\varphi_e \mid \mathbf{s}) - mc^2\right)\left|\varsigma_k^{(2)}(r_e,\varphi_e \mid \mathbf{s},t)\right\rangle +$$
$$i\exp(i\varphi_e)\left(\frac{\partial}{\partial r_e} + \frac{1}{r_e}\frac{\partial}{\partial \varphi_e}\right)\left|\varsigma_k^{(1)}(r_e,\varphi_e \mid \mathbf{s},t)\right\rangle \quad (16b)$$

In Eqs. (16) are replaced the two previous functions in Eqs. (15), i.e, $\varsigma_j(\mathbf{s}_e \mid \mathbf{s},t); j = 1,4$ by the functions $\varsigma_k^{(\lambda)}(r_e,\varphi_e \mid \mathbf{s},t); \lambda = 1,2$. (note that the indices "1" and "4" became now "1" and "2")

Since the electronic potential $V(r_e,\varphi_e|\mathbf{s})$ is time-independent the two eigen-functions in Eqs. (16), are rewritten in the form:

$$\left|\varsigma_k^{(\lambda)}(r_e,\varphi_e \mid \mathbf{s},t)\right\rangle \implies \exp\left((i/\hbar)u_k(\mathbf{s})t\right)\left|\xi_k^{(\lambda)}(r_e,\varphi_e \mid \mathbf{s})\right\rangle; \lambda = \{1,2\} \quad (17)$$

where $u_k(\mathbf{s})$ is still to be determined. Substitution of Eq. (17) in Eqs. (16) yields the two coupled equations for $\xi_k^{(\lambda)}(r_e, \varphi_e | \mathbf{s}); \lambda = 1, 2$:

$$\left(u_k(\mathbf{s}) - mc^2 - V(r_e, \varphi_e | \mathbf{s})\right)\left|\xi_k^{(1)}(r_e, \varphi_e | \mathbf{s})\right\rangle -$$
$$i\exp(-i\varphi_e)\left(\frac{\partial}{\partial r_e} - \frac{1}{r_e}\frac{\partial}{\partial \varphi_e}\right)\left|\xi_k^{(2)}(r_e, \varphi_e | \mathbf{s})\right\rangle = 0 \quad (18a)$$

and

$$\left(u_k(\mathbf{s}) + mc^2 - V(r_e, \varphi_e | \mathbf{s})\right)\left|\xi_k^{(2)}(r_e, \varphi_e | \mathbf{s})\right\rangle -$$
$$i\exp(i\varphi_e)\left(\frac{\partial}{\partial r_e} + \frac{1}{r_e}\frac{\partial}{\partial \varphi_e}\right)\left|\xi_k^{(1)}(r_e, \varphi_e | \mathbf{s})\right\rangle = 0 \quad (18b)$$

To continue we expand $\xi_k^{(\lambda)}(r_e, \varphi_e | \mathbf{s}): \lambda=1,2$ in terms of an ortho-normalized basis set:

$$\left|\xi_k^{(\lambda)}(r_e, \varphi_e | \mathbf{s})\right\rangle = \sum_{n=1}^{N} \varpi_{kn}^{(\lambda)}(\mathbf{s})\left|\eta_n(r_e, \varphi_e | \mathbf{s})\right\rangle; \lambda = 1, 2 \quad (19)$$

Here the summation is expected from 1 to N where N is an arbitrary number, however in the present study we assume that N=2. Substituting Eq. (19) in Eqs. (18) yields the following two corresponding equations::

$$\left(u_k(\mathbf{s}) - mc^2 - V(r_e, \varphi_e \mid \mathbf{s})\right) \sum_{n=1} \varpi_{kn}^{(1)}(\mathbf{s}) \left|\eta_n(r_e, \varphi_e \mid \mathbf{s})\right\rangle$$

(20a)

$$-i\exp(-i\varphi_e)\left(\frac{\partial}{\partial r_e} - \frac{1}{r_e}\frac{\partial}{\partial \varphi_e}\right) \sum_{n=1} \varpi_{kn}^{(2)}(\mathbf{s}) \left|\eta_n(r_e, \varphi_e \mid \mathbf{s})\right\rangle = 0$$

$$\left(u_k(\mathbf{s}) + mc^2 - V(r_e, \varphi_e \mid \mathbf{s})\right) \sum_{n=1} \varpi_{kn}^{(2)}(\mathbf{s}) \left|\eta_n(r_e, \varphi_e \mid \mathbf{s})\right\rangle -$$

(20b)

$$i\exp(i\varphi_e)\left(\frac{\partial}{\partial r_e} + \frac{1}{r_e}\frac{\partial}{\partial \varphi_e}\right) \sum_{n=1} \varpi_{kn}^{(1)}(\mathbf{s}) \left|\eta_n(r_e, \varphi_e \mid \mathbf{s})\right\rangle = 0$$

Multiplying Eqs. (20) by $\eta_j(r_e, \varphi_e \mid \mathbf{s})$; j={1,2} and integrating over ($r_e, \varphi_e$) yields the following algebraic eigen-value equations (which for the two-state molecular system becomes a system of four equations):

$$\begin{pmatrix} u_k - W_{11}^{(1)}(\mathbf{s}) & W_{12}^{(1)}(\mathbf{s}) & D_{11}^{(1)}(\mathbf{s}) & D_{12}^{(1)}(\mathbf{s}) \\ W_{21}^{(1)}(\mathbf{s}) & u_k - W_{22}^{(1)}(\mathbf{s}) & D_{21}^{(1)}(\mathbf{s}) & D_{22}^{(1)}(\mathbf{s}) \\ D_{11}^{(2)}(\mathbf{s}) & D_{12}^{(2)}(\mathbf{s}) & u_k - W_{11}^{(2)}(\mathbf{s}) & W_{12}^{(2)}(\mathbf{s}) \\ D_{21}^{(2)}(\mathbf{s}) & D_{22}^{(2)}(\mathbf{s}) & W_{21}^{(2)}(\mathbf{s}) & u_k - W_{22}^{(2)}(\mathbf{s}) \end{pmatrix} \begin{pmatrix} \varpi_{k1}^{(1)}(\mathbf{s}) \\ \varpi_{k2}^{(1)}(\mathbf{s}) \\ \varpi_{k1}^{(2)}(\mathbf{s}) \\ \varpi_{k2}^{(2)}(\mathbf{s}) \end{pmatrix} == \begin{pmatrix} 0 \\ 0 \\ 0 \\ 0 \end{pmatrix}$$

(21)

Here $W_{jn}^{(\lambda)}(\mathbf{s}); \{\lambda=1,2\}$ stand for the expressions:

$$W_{jn}^{(1)}(s) = \langle \eta_j(r_e, \varphi_e|s) | V(r_e, \varphi_e|s) | \eta_n(r_e, \varphi_e|s) \rangle + mc^2 \delta_{jn} \qquad (22a)$$

$$W_{jn}^{(2)}(s) = \langle \eta_j(r_e, \varphi_e|s) | V(r_e, \varphi_e|s) | \eta_n(r_e, \varphi_e|s) \rangle - mc^2 \delta_{jn} \qquad (22b)$$

and $D_{jn}^{(\lambda)}(s); \{\lambda=1,2\}$ stand for the expressions:

$$D_{jn}^{(1)}(s) = -i \langle \eta_j(r_e, \varphi_e|s) | \exp(-i\varphi_e) \left( \frac{\partial}{\partial r_e} - \frac{1}{r_e} \frac{\partial}{\partial \varphi_e} \right) | \eta_n(r_e, \varphi_e|s) \rangle \qquad (23a)$$

$$D_{jn}^{(2)}(s) = -i \langle \eta_j(r_e, \varphi_e|s) | \exp(i\varphi_e) \left( \frac{\partial}{\partial r_e} + \frac{1}{r_e} \frac{\partial}{\partial \varphi_e} \right) | \eta_n(r_e, \varphi_e|s) \rangle \qquad (23b)$$

## C.2 Implementing the Dirac Eigen-functions in the BOH expansion

At this stage we remind the reader that within the BH formulation the electronic $\varsigma_k(s_e|s)$-eigen-functions are calculated employing Eq.(1) whereas two electronic Dirac eigen-functions $\xi_k^{(\lambda)}(r_e, \varphi_e|s)$ ; $\lambda=1,2$ are formed for each k via the eigen-value equations in Eqs. (18) (see also Eq. (21)). Next, it is realized that Eqs (21) produce four *eigen-values*. However it turns out that only two are relevant as they are close to the BO eigen-values whereas the other two are expected to be far away and therefore will be ignored.

The next step is to construct the BOH-Dirac full wave-function $\Psi(r_e, \varphi_e, s)$ which is expected to satisfy the following (time-independent) Schrödinger equation:

$$\left(-\frac{\hbar^2}{2M}\nabla^2 + H_e(r_e,\varphi_e|\mathbf{s})\right)\Psi(r_e,\varphi_e,\mathbf{s}) = E\Psi(r_e,\varphi_e,\mathbf{s}) \qquad (24)$$

This will be done by expanding $\Psi(r_e,\varphi_e,\mathbf{s})$ in terms of the Dirac eigen-functions namely:

$$\left|\Psi(r_e,\varphi_e,\mathbf{s})\right\rangle = \sum_{k=1}^{2} \left|\xi_k^{(\lambda)}(r_e,\varphi_e|\mathbf{s})\right\rangle \psi_k^{(\lambda)}(\mathbf{s}) \qquad (25)$$

As mentioned earlier for each eigen-value are encountered two Dirac eigen-functions $\xi_k^{(\lambda)}(r_e,\varphi_e|\mathbf{s}); \lambda=1,2$. However according to the above expansion we are capable to apply only one eigen-function at a time. This situation also dictates the relevant expression for the NACTs which are seen to take the form (see Eq. (4)):

$$\boldsymbol{\tau}_{12}^{\lambda} = \left\langle \xi_1^{(\lambda)} \middle| \nabla \xi_2^{(\lambda)} \right\rangle \quad \text{and} \quad \boldsymbol{\tau}_{12}^{(2)\lambda} = \left\langle \xi_1^{(\lambda)} \middle| \nabla^2 \xi_1^{(\lambda)} \right\rangle; \lambda = 1,2 \qquad (26)$$

Thus, substituting Eq. (25) in Eq. (24) and continue the usual procedure for deriving the relevant coupled equations to calculate the nuclear wave functions while employing and Eqs. (26) yields, finally, the two corresponding BOH-Dirac equations:

$$-\frac{\hbar^2}{2M}\nabla^2\psi_1 + \left(u_1 - \frac{\hbar^2}{2M}\boldsymbol{\tau}_{11}^{(2)\lambda} - E\right)\psi_1 - \frac{\hbar^2}{2M}\left(2\boldsymbol{\tau}_{12}^{\lambda}\cdot\nabla + \boldsymbol{\tau}_{12}^{(2)\lambda}\right)\psi_2 = 0$$

$$-\frac{\hbar^2}{2M}\nabla^2\psi_2 + \left(u_2 - \frac{\hbar^2}{2M}\boldsymbol{\tau}_{22}^{(2)\lambda} - E\right)\psi_2 - \frac{\hbar^2}{2M}\left(-2\boldsymbol{\tau}_{12}^{\lambda}\cdot\nabla + \boldsymbol{\tau}_{21}^{(2)\lambda}\right)\psi_1 = 0$$

$$(27)$$

Here, the two eigen-values $u_1$ and $u_2$ are assumed to be the levant approximations, for the expressions:

$$u_k \approx \left\langle \xi_k^{(\lambda)} \middle| \mathbf{H}_e \middle| \xi_k^{(\lambda)} \right\rangle; \quad k=1,2 \tag{28}$$

It is well noticed that the BOH-Dirac equations in Eq. (27) are similar to the ordinary BOH equations given in Eq, (12) except that the PESs and the NACTs which are provided via the Dirac recipe are expected to contain relativistic effects.

## D. Summary

In the present publication is given an extension of the BOH equation to include the Dirac eigen-functions. To start with we derived the Dirac equations (see Appendix I) essentially following Dirac's prescription as given in Ref. 21 but making the necessary changes so that it applies to molecular systems (instead of atoms). The changes are discussed in Sect. C.2 and are presented via the two coupled BOH equations in Eqs. (27).

The motivation for introducing Dirac's eigen-functions is to test to what extent Molecular Fields as derived in Refs. 7 and 8 and given explicitly in Eqs. (5) are affected in case the ordinary NACTs that appear on the r.h.s. of Eqs. (5) as presented explicitly in (6) are replaced by the corresponding Dirac NACTs. If changes are detected this would imply that an electric beam approaching a *ci* and funnel from one cone to the other (see Fig. 1) could be affected by relativistic mass effects.

As a by-product we also derived the BOH-Dirac coupled Schrödinger equations for the nuclei wave functions (see Sect. (C.2)), employing Dirac's eigen-values and eigen-functions.

At this stage we would like to emphasize two facts:

(1) Although the required conditions for having relativistic effects are created when at least three atoms get close enough so that conical intersections are formed, the the particles directly responsible for the relativistic effects are the electrons (and not the nuclei) as they are the ones to be affected most by the external field. Indeed the mathematical treatment shows that the wave equations in Eq. (5) are formed if and only if the Born-Oppenheimer-Huang (BOH) approach is based on the time-dependent eigen-value equation, Eq. (2), and not on Eq. (1).

(2) There is always in the background the question how the relativistic effects can be detected experimentally? The way I see it is based on the assumption that the incoming electric field is affected so that the outgoing field will have eventually different wave lengths and or shifted phases. Next, if indeed incorporating Dirac's relativistic electron theory leads to relativistic effects then these effects may, eventually, be enhanced by increasing the intensity of the external field (thus, in turn, enhancing the signal to be detected).

Finally I would like to make the following comment: The present study centers, as usual, on the existence of the BH NACTs. Whereas the adiabatic BO PESs are well accepted and used routinely in quantum chemistry (which, in turn, establishes the validity of the BOH approach), the NACTs are still considered *bizarre* to quantum chemistry in general and to their *role* within the BH approach in particular. It is difficult to accept that a theory based on two equally important components yields meaningful results when one of the two is ignored. Numerous publications recommend to ignore approaches based on the NACTs mainly because of one irresponsible wretched statement by Mead and Truhlar [23] made about 35 years ago saying: "*Strictly diabatic* bases do not exist for poly-atomic systems…." (The strange idea connected with this statement is that, in fact, no quantum mechanical calculation exists, immaterial for what purpose, which ever produces *strictly numerical results*). This statement should be compared with another statement recently made by Mead: [24] "There may be real cases in which the adiabatic-to-diabatic approach (the one based on the NACTs) is *most* worthwhile". In the Reference section are listed numerous publications which justify this statement: in three of them are compared differential and integral cross sections with experiment [25-27] (based on the BH NACTs) and a few others discussing topological effects to be found within molecular systems [28-50] (again as based on the BH-NACTs)

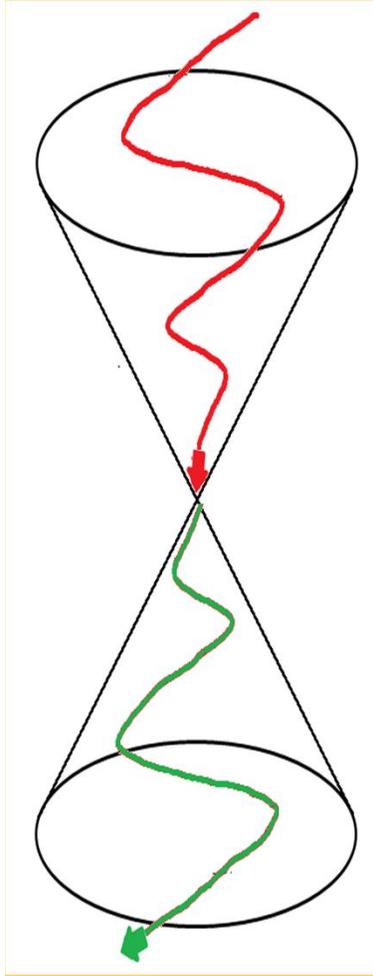

Fig.1 A schematic Figure: An electric beam interacting with two intersecting conical potentials.

# Appendix I

# The Dirac equation for a single electron

The starting point for both treatments is the Schrödinger equation for the total wave-function, $\Psi(\mathbf{s}_e,\mathbf{s})$ given in the form:

$$i\hbar \frac{\partial \Psi(\mathbf{s}_e,\mathbf{s},t)}{\partial t} = \left( -\frac{\hbar^2}{2M}\nabla^2 + \mathbf{H}_e(\mathbf{s}_e|\mathbf{s}) \right) \Psi(\mathbf{s}_e,\mathbf{s},t) \qquad (I.1)$$

where M is the mass of the system, $\nabla$ is the gradient (vector) operator expressed in terms of mass-scaled coordinates, $\mathbf{s}$ and $\mathbf{s}_e$ stand for sets of nuclear and electronic coordinates respectively and $\mathbf{H}_e(\mathbf{s}_e|\mathbf{s})$ is the electronic Hamiltonian which in the present case is assumed to be time-independent. Following BO and BH, [9] the total wave function is written in the form

$$\Psi(\mathbf{s}_e,\mathbf{s},t) = \sum_{j=1}^{N} \eta_j(\mathbf{s}_e|\mathbf{s},t)\psi_j(\mathbf{s}) \qquad (I.2)$$

where $\eta_j(\mathbf{s}_e|\mathbf{s},t)$; j=1,N are the relevant electronic eigen-functions. Within the the BH approach these eigen-functions are formed via the eigen-value time-dependent differential equation:

$$i\hbar \frac{\partial}{\partial t} |\eta_j(\mathbf{s}_e|\mathbf{s},t)\rangle = \mathbf{H}_e(\mathbf{s}_e|\mathbf{s}) |\eta_j(\mathbf{s}_e|\mathbf{s},t)\rangle \qquad (I.3)$$

where **H**$_e$(**s**$_e$|**s**) takes the explicit form:

$$\mathbf{H}_e(\mathbf{s}_e | \mathbf{s}) = -\frac{\hbar^2}{2m_0}\nabla^2 + V(\mathbf{s}_e | \mathbf{s}) \tag{I.4}$$

Here $m_0$ is the rest mass of the electron and $V(\mathbf{s}_e|\mathbf{s})$ is the electronic potential.

To treat the Dirac approach one has to start with the classical expression of the electronic Hamiltonian. For simplicity our study is applied to a single-electron-system for which the Hamiltonian, **H**$_e$(**s**$_e$|**s**), takes the form:

$$\mathbf{H}_e(\mathbf{s}_e | \mathbf{s}) = \frac{\mathbf{P}_e^2}{2m_0} + V(\mathbf{s}_e | \mathbf{s}) \tag{I.5}$$

Here **P**$_e$ is the (classical) non-relativistic momentum of the electron, $m_0$ is its rest mass and $V(\mathbf{s}_e|\mathbf{s})$ stands, as before, for the potential energy of the electron. To continue Eq. (I.5) is modified to include relativistic effects and consequently is written in the form:

$$\mathbf{H}_e(\mathbf{s}_e | \mathbf{s}) = c\left(m^2 c^2 + \mathbf{P}_e^2\right)^{1/2} + V(\mathbf{s}_e | \mathbf{s}) \tag{I.6}$$

where c is the light velocity and m is the corresponding velocity-dependent mass. This equation reduces to Eq. (I.4) (or Eq. (1,5)) when the actual velocity, v, of the electrons is significantly lower than the light velocity, thus when v<<c. Combining Eqs. (I.3) and Eq. (I.6) leads to the corresponding differential eigen-value equation:

$$i\hbar \frac{\partial}{\partial t}\left|\zeta(s_e | s, t)\right\rangle = c\left(m^2 c^2 + P_e^{\,2}\right)^{1/2} \left|\zeta(s_e | s, t)\right\rangle + V(s_e | s)\left|\zeta(s_e | s, t)\right\rangle \tag{I.7}$$

which for v<<c implies that $P_e \ll mc$ and therefore becomes:

$$i\hbar \frac{\partial}{\partial t}\left|\zeta(s_e | s, t)\right\rangle = \left(mc^2 + \frac{1}{2m}P_e^{\,2}\right)\left|\zeta(s_e | s, t)\right\rangle + V(s_e | s)\left|\zeta(s_e | s, t)\right\rangle \tag{I.8}$$

Eq. (I.7) is essentially identical to Eqs. (I.3) because for m~$m_0$ the term $mc^2$ ($\cong m_0 c^2$) becomes a constant and therefore can be ignored.

In his derivation Dirac concentrated on Eq. (I.7) and his idea was to find a way to linearize the relativistic expression $(mc^2+p_e^{\,2})^{(1/2)}$ or $[c(mc^2+p_e^{\,2})^{(1/2)}]$. This could be done by replacing Eq. (I.7) by the equation:

$$i\hbar \frac{\partial}{\partial t}\left|\zeta(s_e | s, t)\right\rangle = c\left(m^2 c^2 + \sum_{j=1}^{3} p_j^{\,2}\right)^{1/2} \left|\zeta(s_e | s, t)\right\rangle + V(s_e | s)\left|\zeta(s_e | s, t)\right\rangle \tag{I.9}$$

where $p_j$ ; j={1,3} are interpreted as operators (in accordance of quantum mechanics). To continue Dirac's approach we refer to the three 4×4 Dirac matrices $\alpha^{(k)}$ ; k={1,3} formed via the three products $\alpha^{(k)} = \rho_1 \sigma_k$; {k=1,3}:

$\alpha^{(1)} = \rho_1 \sigma_1$

$$\boldsymbol{\alpha}^{(1)} = \begin{pmatrix} 0 & 0 & 1 & 0 \\ 0 & 0 & 0 & 1 \\ 1 & 0 & 0 & 0 \\ 0 & 1 & 0 & 0 \end{pmatrix} \begin{pmatrix} 0 & 1 & 0 & 0 \\ 1 & 0 & 0 & 0 \\ 0 & 0 & 0 & 1 \\ 0 & 0 & 1 & 0 \end{pmatrix} = \begin{pmatrix} 0 & 0 & 0 & 1 \\ 0 & 0 & 1 & 0 \\ 0 & 1 & 0 & 0 \\ 1 & 0 & 0 & 0 \end{pmatrix} \qquad \text{(I.10a)}$$

$\alpha^{(2)} = \rho_1 \sigma_2$

$$\boldsymbol{\alpha}^{(2)} = \begin{pmatrix} 0 & 0 & 1 & 0 \\ 0 & 0 & 0 & 1 \\ 1 & 0 & 0 & 0 \\ 0 & 1 & 0 & 0 \end{pmatrix} \begin{pmatrix} 0 & -i & 0 & 0 \\ i & 0 & 0 & 0 \\ 0 & 0 & 0 & -i \\ 0 & 0 & i & 0 \end{pmatrix} = \begin{pmatrix} 0 & 0 & 0 & -i \\ 0 & 0 & i & 0 \\ 0 & -i & 0 & 0 \\ i & 0 & 0 & 0 \end{pmatrix} \qquad \text{(I.10b)}$$

$\alpha^{(3)} = \rho_1 \sigma_3$

$$\boldsymbol{\alpha}^{(3)} = \begin{pmatrix} 0 & 0 & 1 & 0 \\ 0 & 0 & 0 & 1 \\ 1 & 0 & 0 & 0 \\ 0 & 1 & 0 & 0 \end{pmatrix} \begin{pmatrix} 1 & 0 & 0 & 0 \\ 0 & -1 & 0 & 0 \\ 0 & 0 & 1 & 0 \\ 0 & 0 & 0 & -1 \end{pmatrix} = \begin{pmatrix} 0 & 0 & 1 & 0 \\ 0 & 0 & 0 & -1 \\ 1 & 0 & 0 & 0 \\ 0 & -1 & 0 & 0 \end{pmatrix} \qquad \text{(I.10c)}$$

Replacing the function $\zeta(s_e | \mathbf{s}, t)$ in Eq. (I.9) by a column-vector of four functions $\zeta_j (s_e | \mathbf{s}, t)$; j={1,4} and the three momenta $p_j$ ; j={1,3} their corresponding 4×4 matrices leads to the following first order-type Schrödinger equations (in matrix form):

$$-i\hbar \frac{\partial}{\partial t} \begin{pmatrix} \zeta_1(\mathbf{s}_e|\mathbf{s},t) \\ \zeta_2(\mathbf{s}_e|\mathbf{s},t) \\ \zeta_3(\mathbf{s}_e|\mathbf{s},t) \\ \zeta_4(\mathbf{s}_e|\mathbf{s},t) \end{pmatrix} = ic\hbar \left[ \begin{pmatrix} 0 & 0 & 0 & 1 \\ 0 & 0 & 1 & 0 \\ 0 & 1 & 0 & 0 \\ 1 & 0 & 0 & 0 \end{pmatrix} \frac{\partial}{\partial s_{ex}} + \begin{pmatrix} 0 & 0 & 0 & -i \\ 0 & 0 & i & 0 \\ 0 & -i & 0 & 0 \\ i & 0 & 0 & 0 \end{pmatrix} \frac{\partial}{\partial s_{ey}} + \right.$$

(I.11)

$$\left. \begin{pmatrix} 0 & 0 & 1 & 0 \\ 0 & 0 & 0 & -1 \\ 1 & 0 & 0 & 0 \\ 0 & -1 & 0 & 0 \end{pmatrix} \frac{\partial}{\partial s_{ez}} + \left( V(\mathbf{s}_e|\mathbf{s}) + \beta mc^2 \right) \right] \begin{pmatrix} \zeta_1(\mathbf{s}_e|\mathbf{s},t) \\ \zeta_2(\mathbf{s}_e|\mathbf{s},t) \\ \zeta_3(\mathbf{s}_e|\mathbf{s},t) \\ \zeta_4(\mathbf{s}_e|\mathbf{s},t) \end{pmatrix}$$

Eqs (I.11) can be written in the following compact way:

$$-i\hbar \frac{\partial}{\partial t} \left| \zeta_j(\mathbf{s}_e | \mathbf{s}, t) \right\rangle = \left( V(\mathbf{s}_e | \mathbf{s}) + \beta mc^2 \right) \left| \zeta_j(\mathbf{s}_e | \mathbf{s}, t) \right\rangle +$$
$$ic\hbar \sum_{k=1}^{3} \sum_{i=1}^{4} \alpha_{ji}^{(k)} \frac{\partial}{\partial s_{ek}} \left| \zeta_i(\mathbf{s}_e | \mathbf{s}, t) \right\rangle; \quad j = \{1,4\}$$

(I.12)

Here (and in Eq. (I.11)) β is the metric in SU(2), namely equal to +1 for the two upper diagonal terms and to -1 for the two lower one.

Dirac studied the energy levels of the Hydrogen atom while employing two out of the four equations given in Section C.1, namely, Eqs. (14a) and (14d). We consider the same two coupled equations (I.9a) and (I.9d) (and ignore the third term in each one of them thus leading to a planar electronic system)

$$\left(-i\hbar\frac{\partial}{\partial t}+V(\mathbf{s}_e\,|\,\mathbf{s})-mc^2\right)|\zeta_1(\mathbf{s}_e\,|\,\mathbf{s},t)\rangle=$$

$$i c\hbar\left(\frac{\partial}{\partial s_{ex}}-i\frac{\partial}{\partial s_{ey}}\right)|\zeta_4(\mathbf{s}_e\,|\,\mathbf{s},t)\rangle \qquad\text{(I.13a)}$$

$$\left(-i\hbar\frac{\partial}{\partial t}+V(\mathbf{s}_e\,|\,\mathbf{s})+mc^2\right)|\zeta_4(\mathbf{s}_e\,|\,\mathbf{s},t)\rangle=$$

$$i c\hbar\left(\frac{\partial}{\partial s_{ex}}+i\frac{\partial}{\partial s_{ey}}\right)|\zeta_1(\mathbf{s}_e\,|\,\mathbf{s},t)\rangle \qquad\text{(I.13b)}$$

In what follows we consider the planar case and therefore are applied polar coordinates, $s_x = r_e \cos\varphi_e$; $s_y = r_e \sin\varphi_e$, so that:

$$\frac{\partial}{\partial s_{ex}} = \cos\varphi_e \frac{\partial}{\partial r_e} - \sin\varphi \frac{1}{r_e}\frac{\partial}{\partial \varphi_e} \qquad\text{(I.14a)}$$

$$\frac{\partial}{\partial s_{ey}} = \sin\varphi_e \frac{\partial}{\partial r_e} + \cos\varphi \frac{1}{r_e}\frac{\partial}{\partial \varphi_e} \qquad\text{(I.14b)}$$

Consequently, the corresponding two differential operators in Eqs. (I.13) take the form:

$$\frac{\partial}{\partial x} - i\frac{\partial}{\partial y} = \exp(-i\varphi)\left(\frac{\partial}{\partial r} - i\frac{1}{r}\frac{\partial}{\partial \varphi}\right) \qquad\text{(I.15a)}$$

$$\frac{\partial}{\partial x} + i\frac{\partial}{\partial y} = \exp(i\varphi)\left(\frac{\partial}{\partial r} + i\frac{1}{r}\frac{\partial}{\partial \varphi}\right) \tag{I.15b}$$

and two coupled Dirac Equations (to be used in the main text) become:

$$\left(-i\hbar\frac{\partial}{\partial t} + V(r_e, \varphi_e \mid \mathbf{s}) - mc^2\right)\left|\varsigma_1(r_e, \varphi_e \mid \mathbf{s}, t)\right\rangle = $$
$$ic\hbar\exp(-i\varphi_e)\left(\frac{\partial}{\partial r_e} - \frac{1}{r_e}\frac{\partial}{\partial \varphi_e}\right)\left|\varsigma_4(r_e, \varphi_e \mid \mathbf{s}, t)\right\rangle \tag{I.16a}$$

$$\left(-i\hbar\frac{\partial}{\partial t} + V(r_e, \varphi_e \mid \mathbf{s}) + mc^2\right)\left|\varsigma_4(r_e, \varphi_e \mid \mathbf{s}, t)\right\rangle = $$
$$ic\hbar\exp(i\varphi_e)\left(\frac{\partial}{\partial r_e} + \frac{1}{r_e}\frac{\partial}{\partial \varphi_e}\right)\left|\varsigma_1(r_e, \varphi_e \mid \mathbf{s}, t)\right\rangle \tag{I.16b}$$